# Electrochemical and mechanical behaviors of dissimilar friction stir welding between 5086 and 6061 aluminum alloy

Zhitong Chen[*], Shengxi Li, Lloyd H. Hihara[*]

Hawaii Corrosion Laboratory, Department of Mechanical Engineering, University of Hawaii at Manoa, Honolulu, HI 96822, USA

*Abstract*

The mechanical properties and corrosion behavior of friction stir welded AA5086 and AA6061 Al alloys were investigated. Micro-hardness measurements and tensile tests showed that the heat-affected zone (HAZ) in AA6061 had minimum hardness value (i.e., 88 HV) and served as failure site in the dissimilar weld. Corrosion testing revealed that the minimum value of $I_{corr}$ appeared in the HAZ 5086 (0.54 µA/cm$^2$) and HAZ 5086 was most resistant to corrosion. The AA 5086 side of the weld showed better corrosion resistance than the AA 6061 side.

---

[*] Corresponding authors:
E-mail addresses: jeadonchen@gmail.com, hihara@hawaii.edu

*1 Introduction*

Friction stir welding (FSW) has emerged as a new solid state technique for joining metallic materials [1], especially Al alloys [2, 3]. The FSW joining is achieved by a non-consumable rotating tool with specially designed pin and shoulder, which inserts into the abutting edges of metal sheets and traverses along the line of the joint [4, 5]. FSW can be considered as a hot-working process where a large amount of deformation is imparted to the workpieces through the rotating pin and shoulder. This process generally results in three different welding zones: the nugget zone (NZ), the thermomechanically-affected zone (TMAZ), and the heat-affected zone (HAZ) [6-11]. The NZ experiences the highest temperature and the highest plastic deformation as compared to other regions, and therefore usually consists of fine equiaxed grains due to full dynamic recrystallization. The HAZ is only affected by heat without plastic deformation, while TMAZ adjacent to the NZ is plastically deformed and heated. When compared to traditional welding technologies, FSW significantly eliminates large distortions, solidification cracking, oxidation, high porosity, etc. [12]. In addition, many metallurgical reactions between dissimilar materials at elevated temperatures can be avoided by using FSW. As an example, FSW is a prospective welding technique for joining dissimilar Al alloys having incompatibilities [13].

The joining of dissimilar Al alloys using FSW has been extensively investigated, such as the effect of welding speed and fixed location of base Al alloys on microstructures, hardness distributions, and tensile properties of the welded joints [14-17]. Aval et al. studied the effect of tool geometry on mechanical and microstructural properties of FSW AA5086-AA6061 [18]. Ilangovan et al. investigated microstructure and tensile properties of FSW AA6061-AA5086 [19]. Despite extensive studies on dissimilar Al alloy FSW joints, there is a lack of understanding of their corrosion behavior.

In the present work, the corrosion behavior of AA5086-AA6061 joints by FSW was examined using electrochemical and immersion tests in three different solutions, i.e., 3.15 wt% NaCl, ASTM seawater, and 0.5 M $Na_2SO_4$. The electrochemical tests were conducted using sectioned samples representing each different zone, while immersion tests were carried out using the as-fabricated joints after grinding.

*2 Experimental*

AA5086-H32 (0.4% Si, 0.5% Fe, 0.1% Cu, 0.35% Mn, 4.0% Mg, 0.15% Cr, 0.25% Zn, 0.15% Ti, balance Al) and AA 6061-T6511 (0.4% Si, 0.7% Fe, 0.4% Cu, 0.15% Mn, 1.2% Mg, 0.35% Cr, 0.25% Zn, 0.15% Ti, balance Al) plates were friction stir welded vertical to the rolling direction with a travel speed, a rotational speed, and a shoulder diameter of 20 mm/min, 1000 rpm and 25 mm, respectively. The friction stir pin had a diameter of 8 mm and height of 6.35 mm. A simultaneous rotation and translation motion of the FSW tool generates the formation of an asymmetric weld. When the tool rotates in the direction of its translation, it refers to the advancing side (AS). When rotation and translation of the tool are in the opposite direction, it refers to the retreating side (RS) (Fig. 1). Fig. 2 shows the schematic view of the cross-section of the FSW joint of AA5086-AA6061.

Vickers microhardness testing (Wilson Rockwell, R5000) was performed across the welds using a 50 g load. The tensile tests were performed with a crosshead speed of 3 mm/min using an Instron-5500R testing machine. Two types of tensile specimens were machined from the FSW joints: parallel (longitudinal) and normal (transverse), as depicted in Fig. 1. An axial extensometer with a 25 mm gage length was attached to the specimens at the gauge section. The strain analysis of each specimen was achieved using an ASAME automatic strain measuring system. Three tensile specimens cut from the same joint were evaluated.

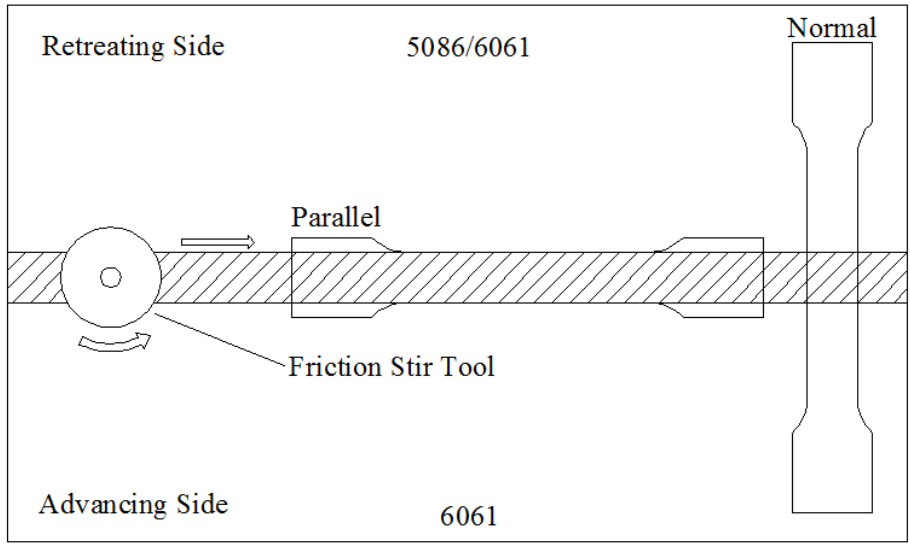

Fig. 1. Schematic showing the FSW joining process and the locations where tensile specimens were obtained.

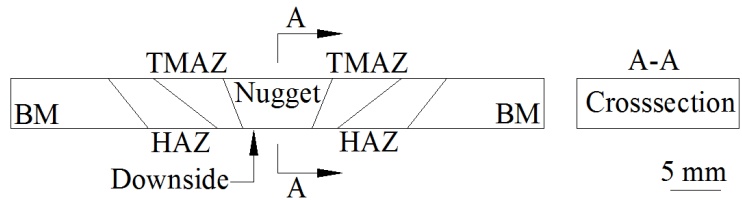

Fig. 2. Schematic view of the cross-section of the FSW joint of AA5086-AA6061.

Polarization experiments were conducted using samples sectioned from different regions, i.e., NZ, TMAZ, HAZ and base metal (BM), using a low-speed diamond saw (Buehler). The samples were mounted in epoxy resin, polished to a 0.05 μm mirror-finish, and immersed in high purity water (18 MΩ cm) prior to polarization experiments in 3.15 wt.% NaCl at 30°C. The solution was deaerated with high-purity nitrogen (> 99.999%). Potentiodynamic polarization experiments were conducted with a PARSTAT 2273 potentiostat. The working electrodes were kept in the open-circuit condition for 1 h prior to conducting the potentiodynamic scan at a rate of 1 mV/s. A saturated calomel electrode (SCE) was used as the reference electrode and a platinum mesh was used as the counter electrode. To minimize contamination, the reference electrode was kept in a separate cell connected via a Luggin probe. Anodic and catholic sweeps were measured separately on freshly-prepared samples. Polarization experiments for each typical zone were performed using at least three samples to verify reproducibility. To generate final polarization diagrams, the mean values of the logarithm of the current density were plotted as a function of potential.

For immersion tests, FSW specimens with dimensions of 70 mm × 25 mm × 4 mm were prepared. The specimens were degreased in acetone, ultrasonically cleaned in deionized water, dried, and weighed to obtain the initial mass. The specimens in triplicates were mounted to acrylic holders and placed in 250 ml beakers. A total of 24 beakers were placed in an aquarium chamber maintained at 30 ºC. Approximately 200 ml of 3.15 wt.% NaCl, 0.5 M $Na_2SO_4$, and ASTM seawater solutions was poured into the beaker so that the specimens were completely immersed. The beakers were partially covered using plastic Petri dishes to minimize evaporation and maintain aerated conditions. After 90 and 120 days of immersion, the specimens were retrieved, dried in a dry box (1% RH), and characterized using scanning electron microscopy (SEM) and Raman spectroscopy. Then, the specimens were cleaned in a solution of phosphoric acid ($H_3PO_4$) and chromium trioxide ($CrO_3$) at 90 ºC for 10 min, as described in ASTM G01-03. Finally, the cleaned specimens were weighed to calculate the weight loss.

## 3 Results and Discussion

### 3.1 Mechanical properties

Fig. 3 shows the microhardness across the top surface of FSW AA5086-AA6061, which does not have a typical "W"-shaped hardness distribution characteristic. A minimum hardness of 88 HV is obtained in the HAZ on the AA6061 side, which suggests that the tensile specimens are prone to fracture in this zone. The hardness distribution on the AA6061 side of the AA5086-AA6061 FSW joint is similar to that of the AA6061-AA6061 FSW joint [20]. However, the hardness distribution on the AA5086 side of the AA5086-AA6061 FSW joint does not resemble that of the AA5086-AA5086 FSW joint [21, 22]. For the AA5086 side, the hardness of BM was higher than that of the TMAZ/HAZ. In the NZ, the hardness increased from the AA6061 side to the AA5086 side.

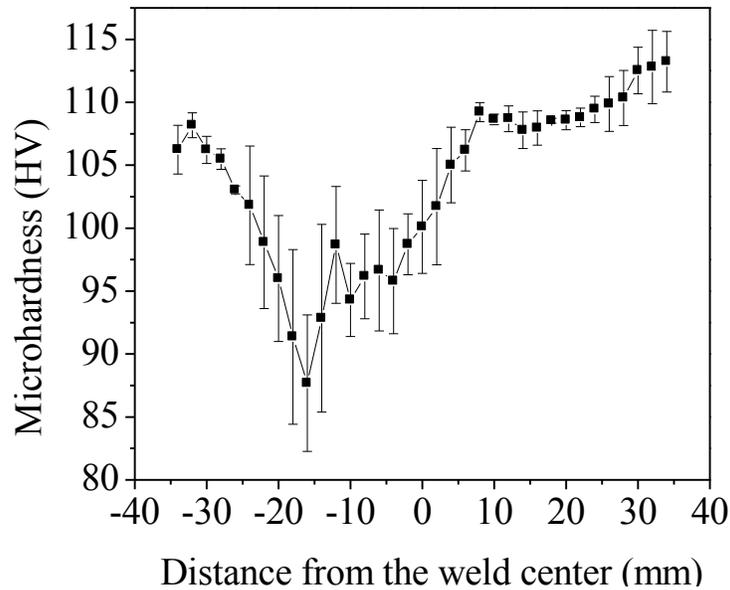

Fig. 3. Microhardness distribution across the top surface of AA5086-AA6061 FSW joint measured with a 2 mm step.

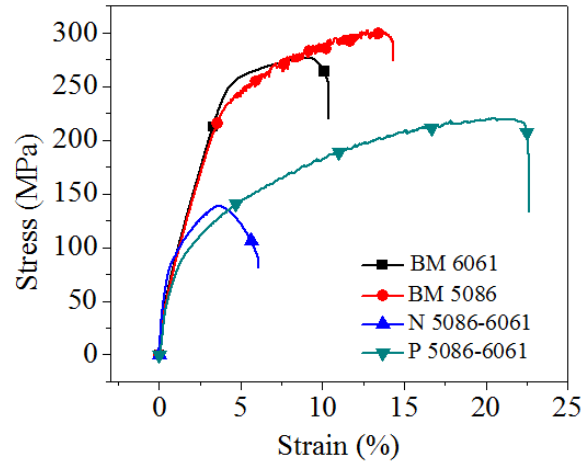

Fig. 4. Tensile test results of the AA5086-AA6061 FSW specimens as compared to those of base metals AA5086 and AA6061. N: normal, P: parallel.

Fig. 4 shows the tensile test results of specimens from the AA5086-AA6061 FSW joints, as compared to those of the base metals. Both FSW specimens (N and P) had lower tensile and yield strength than the base metals, indicating the negative effect of FSW process on the mechanical properties of the joints. In terms of ductility, the parallel FSW tensile specimen had a significantly increased ductility as compared to the base metals, while the normal FSW tensile specimen showed a much lower ductility than the base metals.

From hardness results, we know that the hardness values of the NZ were lower than that of the BM, possibly explaining why the longitudinal tensile specimens of FSW joints decreased in both tensile and yield strength. The transverse tensile specimens contained all four zones (i.e., BM, HAZ, TMAZ and NZ). The observed ductility was measured as average strain over the gage length, including the various zones that have different resistances to deformation due to differences in grain size and precipitate distribution. When a tensile load was applied to the joint, failure occurred in the relatively weak regions of the joint [23]. The fracture of the transverse direction tensile specimens occurred in the HAZ.

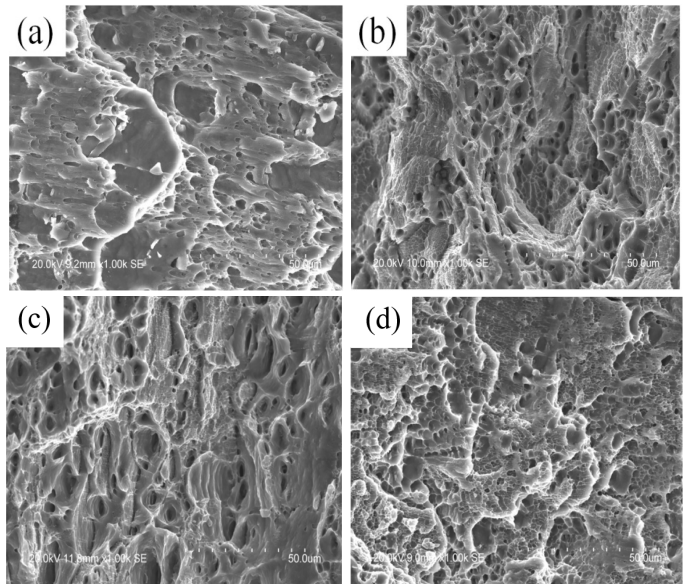

Fig. 5. SEM fractographs of BM and FSW specimens: (a) BM AA5086, (b) BM AA6061, (c) N AA5086-AA6061 FSW (failure occurred in HAZ on the AA6061 Side), and (d) P AA5086-AA6061 FSW (failure occurred in weld zone).

Most of the specimens had a 45° angle shear fractures along the tensile axis with the exception of BM AA6061 and N AA5086-AA6061 FSW specimens. Obvious necking/plastic deformation was observed on the FSW tensile specimens but not on BM AA5086 and BM AA6061 specimens. Fig. 5 presents SEM fractographs of the BM and FSW specimens. The fractographs reveal dimple fracture patterns with teared edges full of micropores. The dimples were of various sizes and shapes. The fracture surface of BM AA6061 has deeper dimples and thinner teared edges than the BM AA5086 specimen, which agrees with the fact that the BM AA5086 specimen exhibited better mechanical properties than the BM AA6061 specimen (Fig. 4). Fig. 5 also shows that the fracture surface of the N AA5086-AA6061 FSW specimen has much deeper dimples and thinner teared edges than the P AA5086-AA6061 FSW specimen, indicating that the P AA5086-AA6061 FSW specimen had better mechanical properties than the N AA5086-AA6061 FSW specimen.

*3.2 Electrochemical measurements*

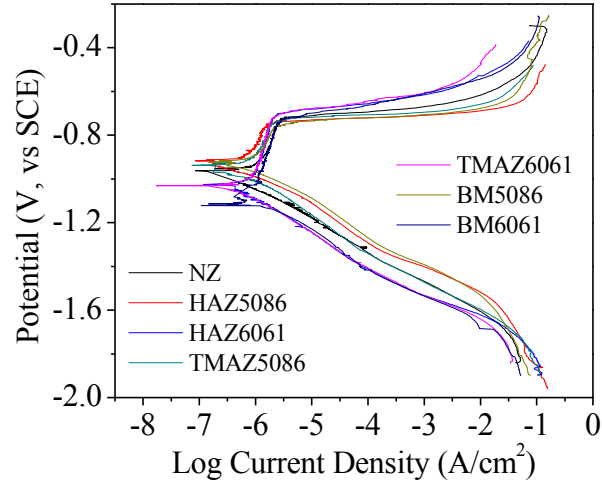

Fig. 6. Polarization curves of samples from different zones in the AA5086-AA6061 FSW joint in deaerated 3.15 wt.% NaCl solutions.

Table 1. $E_{corr}$, $I_{corr}$ and $E_{pit}$ values obtained from the polarization results of samples from different zones in the AA5086-AA6061 FSW joint in deaerated 3.15 wt.% NaCl solutions.

| Weld | $E_{corr}$ (mV, vs SCE) | $E_{pit}$ (mV, vs SCE) | $I_{corr}$ (µA/cm$^2$) | S.D. of $I_{corr}$ |
|---|---|---|---|---|
| HAZ 5086 | -917 | -739 | 0.54 | 0.11 |
| HAZ 6061 | -1025 | -705 | 1.17 | 0.46 |
| TMAZ 5086 | -937 | -734 | 0.93 | 0.41 |
| TMAZ 6061 | -1031 | -714 | 1.25 | 0.01 |
| BM 5086 | -953 | -767 | 1.70 | 0.07 |
| BM 6061 | -1114 | -724 | 2.44 | 0.71 |
| NZ | -962 | -718 | 1.21 | 0.33 |

Fig. 6 shows polarization curves of samples from different zones in the AA5086-AA6061 FSW joint in deaerated 3.15 wt.% NaCl solutions. The passive regions of FSW AA5086-AA6061 coupons in AA6061 side (BM-AA6061, HAZ-AA6061, and TMAZ-AA6061) were larger than those of the AA5086 side. $E_{corr}$, $I_{corr}$ and $E_{pit}$ values of the different positions of FSW AA5086-AA6061 in deaerated 3.15 wt% NaCl solutions are shown in Table 1. The $E_{corr}$ values of downside (-944 mV$_{SCE}$) shifted to more anodic potentials from the BM-AA5086 (-953 mV$_{SCE}$) and BM-AA6061 (-1114 mV$_{SCE}$), while the $E_{corr}$ value of the NZ was between the BM-AA5086 and BM-AA6061. The $E_{corr}$ values of the HAZ-5086 (-917 mV$_{SCE}$) and the TMAZ-AA5086 (-937 mV$_{SCE}$) shifted to more positive potentials from those of the BM-AA5086 (-953 mV$_{SCE}$).

The same results occurred on the AA6061 side. The $I_{corr}$ values decreased compared to the BM, as shown in Table 1. The $E_{pit}$ values of downside (-749 mV$_{SCE}$) shifted to more negative potentials compared to those of the BM-AA5086 (-737 mV$_{SCE}$) and BM-AA6061 (-724 mV$_{SCE}$), while those of the NZ (-718 mV$_{SCE}$) shifted to more positive values. For the AA6061 side, the $E_{pit}$ values of the HAZ-AA6061 (-705 mV$_{SCE}$) and the TMAZ-AA6061 (-714 mV$_{SCE}$) shifted to more positive values compared to those from the BM-AA6061 (-724 mV$_{SCE}$). The FSW improved corrosion resistance of the alloy mixtures compared to the AA5086 and AA6061 BM.

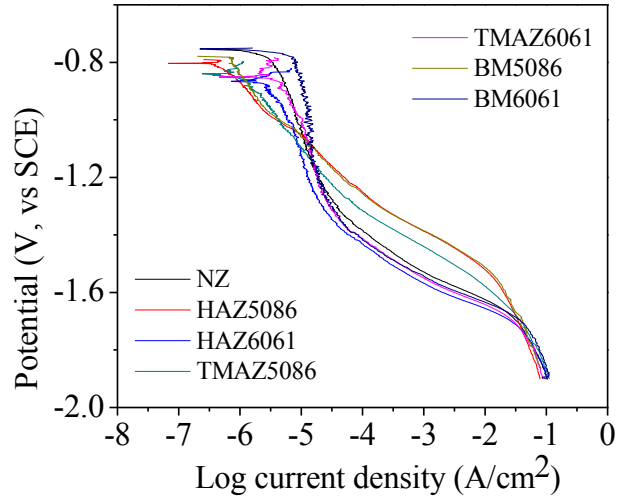

Fig. 7. Cathodic polarization diagram of different weld zones of FSW AA5086-AA6061 in aerated 3.15 wt% NaCl solutions.

Table 2. $E_{corr}$ and $I_{corr}$ values of the different weld zones of FSW AA5086-AA6061 in aerated 3.15 wt% NaCl solutions and standard deviation of $I_{corr}$

| Weld | $E_{corr}$ (mV, vs SCE) | $I_{corr}$ (μA/cm$^2$) | S.D. of $I_{corr}$ |
|---|---|---|---|
| HAZ 5086 | -790 | 0.53 | 0.16 |
| HAZ 6061 | -820 | 0.56 | 0.17 |
| TMAZ 5086 | -797 | 0.87 | 0.15 |
| TMAZ 6061 | -783 | 4.73 | 2.09 |
| BM 5086 | -779 | 1.27 | 0.01 |
| BM 6061 | -750 | 8.57 | 1.24 |
| NZ | -753 | 1.44 | 0.39 |

Fig. 7 shows a cathodic polarization diagram of different weld zones of FSW AA5086-AA6061 in aerated 3.15 wt% NaCl solutions. In aerated 3.15 wt% NaCl the regions of HAZ-AA6061, TMAZ-AA6061, BM-AA6061, and upside show behavior of diffusion-limited oxygen reduction, while the regions of HAZ-AA5086, TMAZ-AA5086, and BM-AA5086 show Tafel behavior. Table 2 shows $E_{corr}$ and $I_{corr}$ values of the different weld zones of FSW AA5086-AA6061 in aerated 3.15 wt% NaCl solutions. The $E_{corr}$ values of different weld zones of FSW AA5086-AA6061 shifted to more negative values than those of BM-AA5086 (-779 mV$_{SCE}$) and BM-AA6061 (-753 mV$_{SCE}$). Compared to the BM-AA6061 (8.57 µA/cm$^2$), the $I_{corr}$ values of the HAZ-AA6061 (0.56 µA/cm$^2$), TMAZ-AA6061 (4.73 µA/cm$^2$), and NZ (1.44 µA/cm$^2$) decreased. Also, the $I_{corr}$ values of HAZ-AA5086 (0.53 µA/cm$^2$), and TMAZ (0.87 µA/cm$^2$) decreased compared to those of BM-AA5086 (1.27 µA/cm$^2$).

### 3.3 Corrosion attack

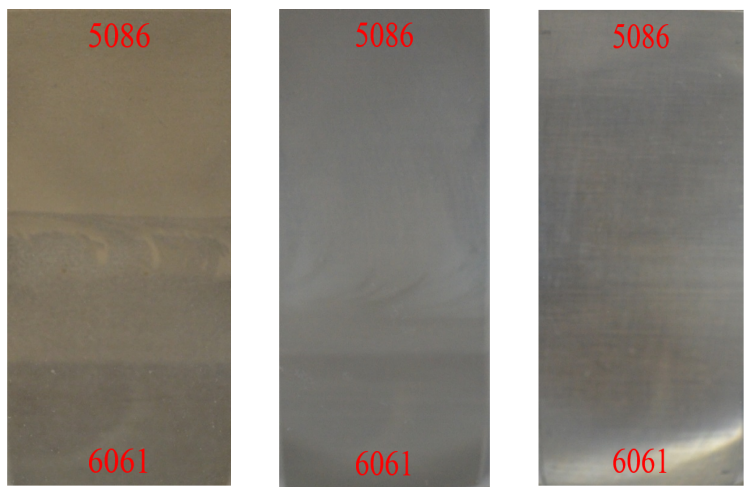

Fig. 8. FSW 5086-6061 samples immersed in (a) 3.15 wt.% NaCl, (b) ASTM seawater, and (c) 0.5 M Na$_2$SO$_4$ solution for 90 days.

Fig. 8 shows FSW AA5086-AA6061 coupons in 3.15 wt.% NaCl, ASTM seawater, and 0.5 M Na$_2$SO$_4$ solutions after 90 days. In 3.15 wt.% NaCl and ASTM seawater, the BM 6061 side was much darker than the BM 5086 side, while the welded regions of those coupons were more complex. The NZ and adjacent regions contain a mixture of AA5086 and AA6061 due to solid-state mixing during the FSW process. Two colors indicate the different corrosion rates of AA5086 and AA6061. As shown in Fig. 8, there was almost no corrosion product on the surface of FSW AA5086-AA6061 coupons after 90 days immersion in 0.5 M $Na_2SO_4$ solution. After 120 days immersion in 0.5 M $Na_2SO_4$ solution, there was much more corrosion product on the BM 6061 side than on the BM 5086 side. There was almost no corrosion product on the BM 5086 side after immersion for 120 days in 0.5 M $Na_2SO_4$ solution.

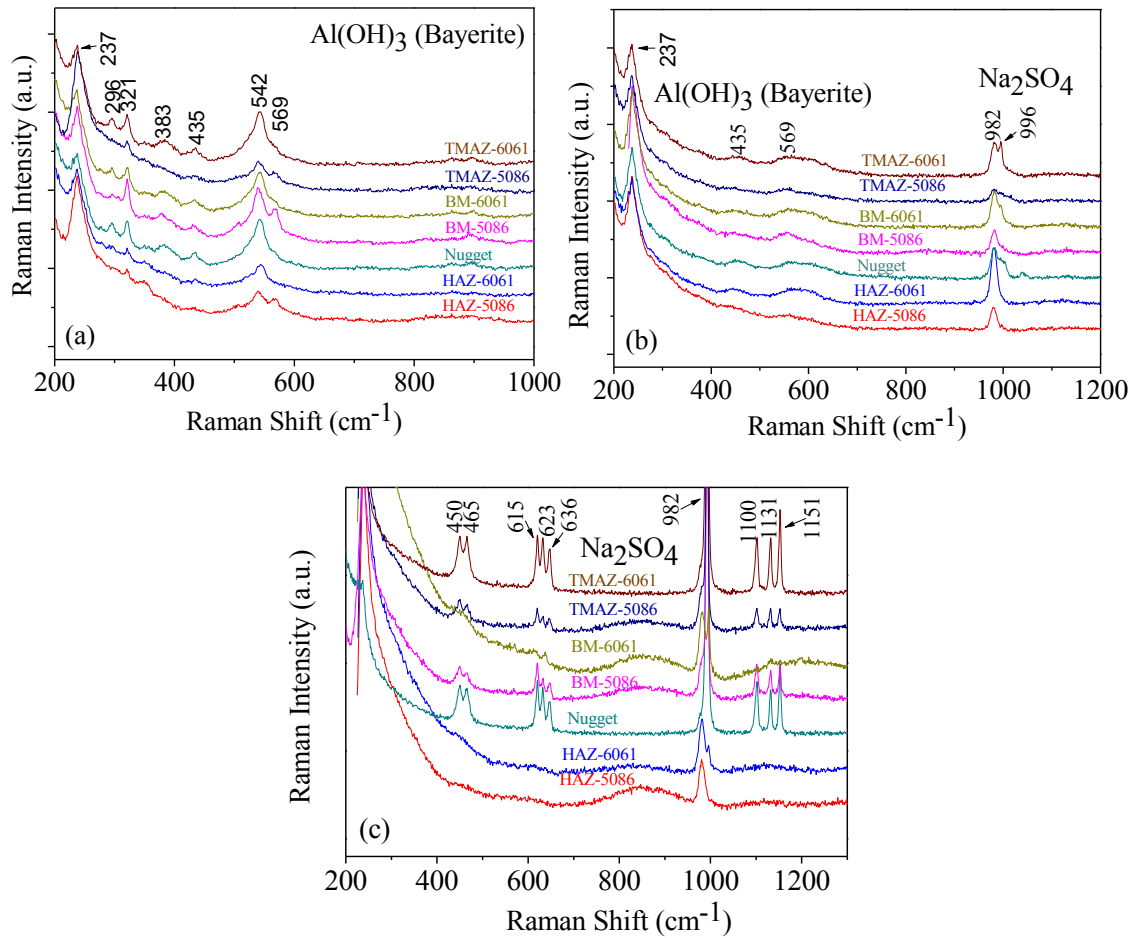

Fig. 9. Raman spectroscopy of distinct zones of FSW AA5086-AA6061 after 90 days immersion in: (a) 3.15 wt% NaCl, (b) ASTM seawater, and (c) 0.5 M $Na_2SO_4$ solution.
12

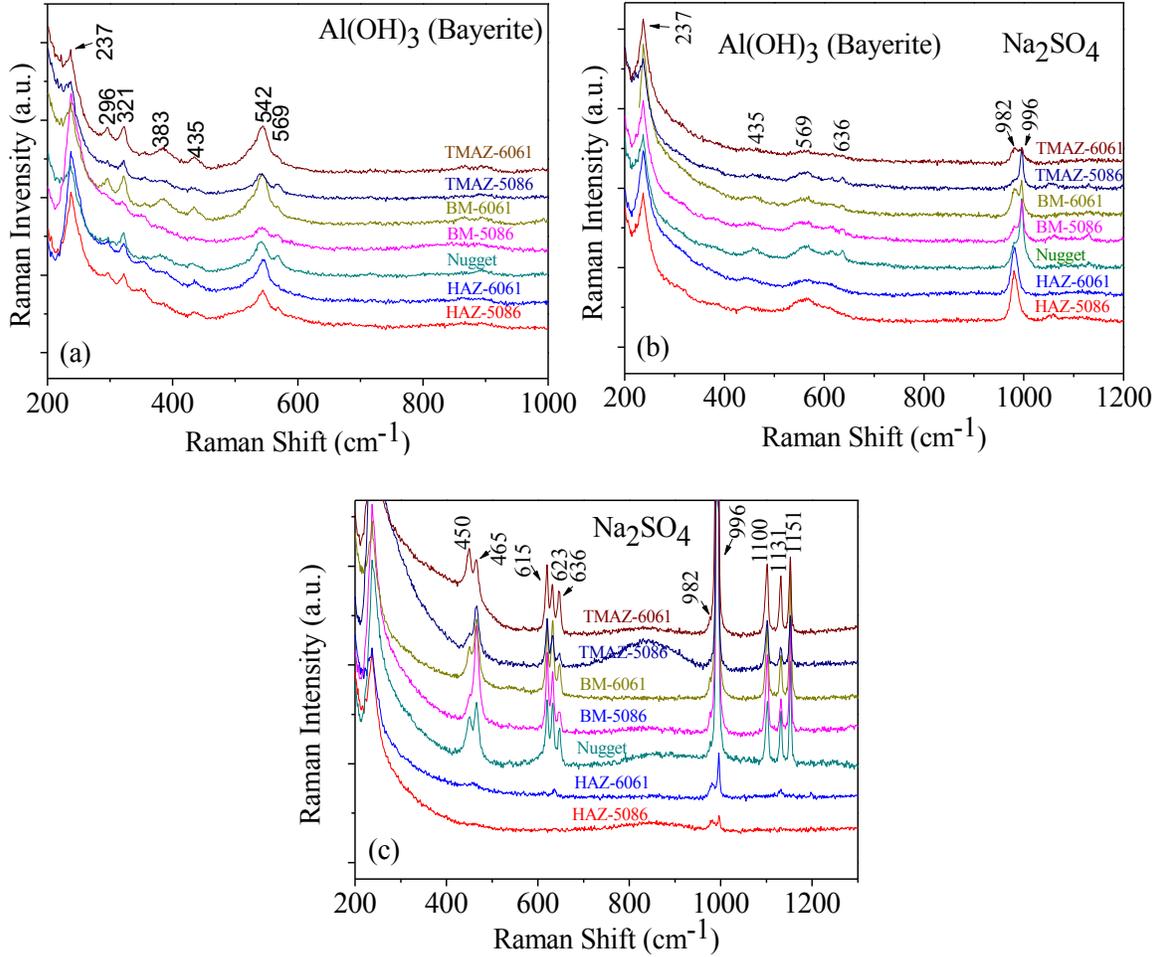

Fig. 10. Raman spectroscopy of distinct zones of FSW AA5086-AA6061 after 120 days immersion in: (a) 3.15 wt% NaCl, (b) ASTM seawater, and (c) 0.5 M $Na_2SO_4$ solution.

Fig. 9 and 10 show Raman spectra of distinct zones of FSW AA5086-AA6061 after 90 and 120 days immersed in 3.15 wt% NaCl, ASTM seawater, and 0.5 M $Na_2SO_4$ solution. We characterized both the AA5086 and AA6061 sides for NZ, TMAZ, BM, and HAZ via Raman spectroscopy. In Fig. 9 and Fig. 10a, characteristic bands of $Al(OH)_3$ (237, 296, 321, 383, 435, 542, and 569 cm$^{-1}$) were observed for distinct regions. TMAZ-AA5086, BM-AA5086, and HAZ-AA5086 had lower Raman intensity than the 6061 side, because there was less corrosion product on the AA5086 side, as shown in Fig. 8. In Fig. 9b and Fig. 10b, characteristic bands of $Al(OH)_3$ (237, 435, 569, and 636 cm$^{-1}$) and $Na_2SO_4$ (982, and 996 cm$^{-1}$) were observed for all regions of FSW AA5086-AA6061 immersed in 0.5 M Na$_2$SO$_4$ solution. In Fig. 9c and Fig. 10c, characteristic bands of Na$_2$SO$_4$ were observed in TMAZ-AA6061, TMAZ-AA5086, BM-AA6061, BM-AA5086, and NZ. However, only bands of 982 and 996 cm$^{-1}$ were found in the regions of HAZ-AA6061 and HAZ-AA5086. Raman data of Al(OH)$_3$ bands for FSW AA5086-AA6061 coupons immersed in 0.5 M Na$_2$SO$_4$ solution were still missing, probably because there was insufficient corrosion product present. Notice that the strong Raman bands at 237 cm$^{-1}$ might be partially from system noise [24-26].

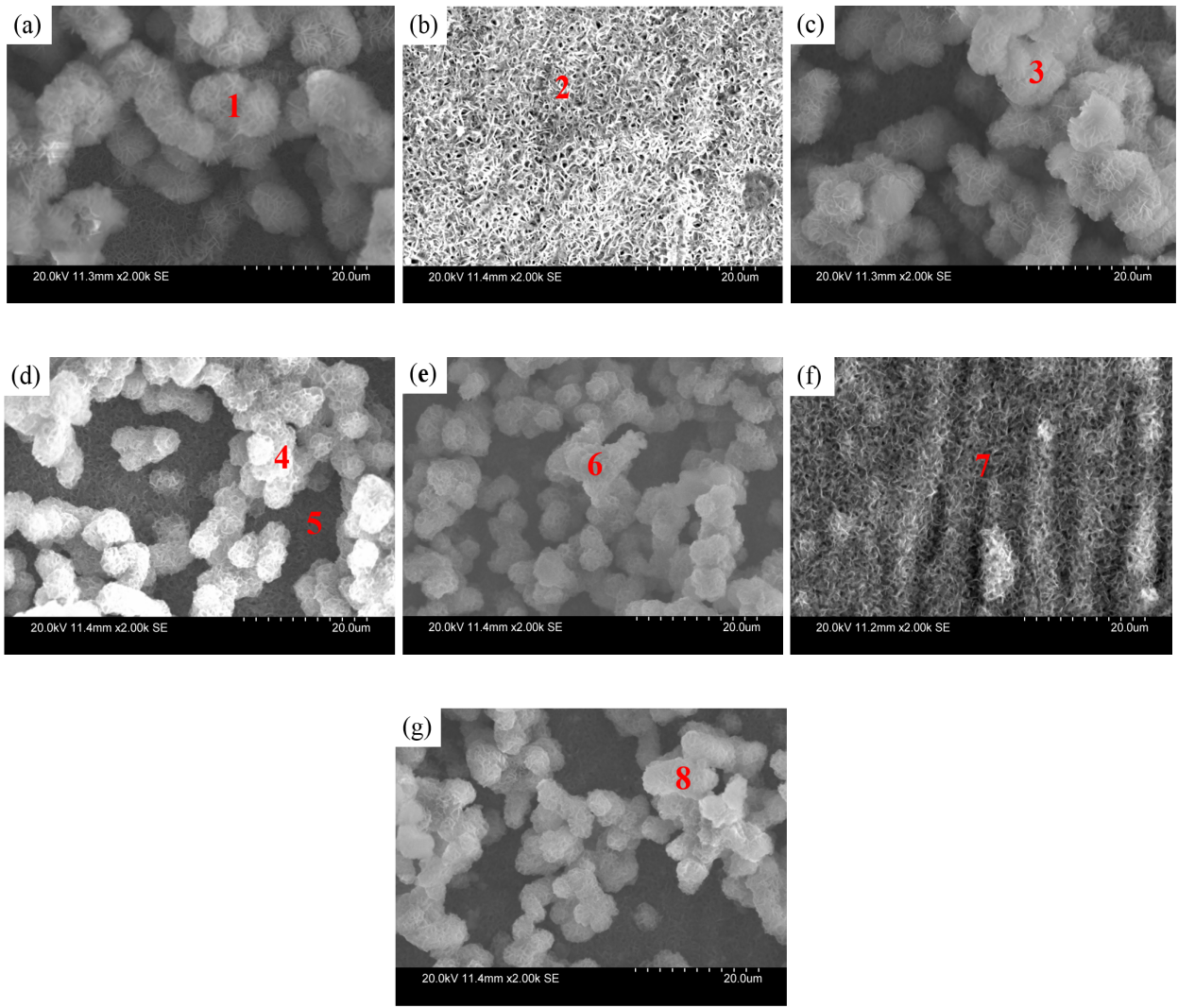

Fig. 11. Different zones (a) BM 5086 (b) HAZ 5086 (c) TMAZ 5086 (d) Nugget (e) TMAZ 6061 (f) HAZ 6061 (g) BM 6061 of FSW AA5086-AA6061 after 120 days immersion in ASTM seawater.

Table 3. EDXA quantification results of six points in Fig. 11.

| Element [at. %] | S | O | Na | Al | Cl | Mg |
|---|---|---|---|---|---|---|
| Spectrum 1 | 4.10 | 60.79 | 3.79 | 12.10 | 9.18 | 10.04 |
| Spectrum 2 | 2.58 | 59.00 | 0.44 | 31.62 | 2.83 | 3.53 |
| Spectrum 3 | 3.79 | 63.24 | 4.11 | 10.86 | 7.54 | 10.46 |
| Spectrum 4 | 2.59 | 66.89 | 4.27 | 10.84 | 5.80 | 9.61 |
| Spectrum 5 | 4.30 | 52.01 | 0.04 | 23.83 | 12.56 | 7.26 |
| Spectrum 6 | 2.37 | 68.57 | 2.99 | 9.30 | 6.14 | 10.63 |
| Spectrum 7 | 1.14 | 62.25 | 3.33 | 20.00 | 3.70 | 9.58 |
| Spectrum 8 | 2.38 | 69.17 | 0.88 | 10.67 | 7.61 | 9.29 |

Fig. 11 shows the SEM of seven regions of FSW AA5086-AA6061 after 120 days immersion in ASTM seawater. There was more corrosion product on the surface of the BM 5086, TMAZ 5086, Nugget, BM 6061, and TMAZ 6061 than on the HAZ 5086 and HAZ 5086. EDXA quantification results in Table 3 reveal that the corrosion product was also high in oxygen and aluminum, which agrees with the Raman results Fig. 10b. The main corrosion product on the seven regions was $Al(OH)_3$. The low corrosion on the downside of samples could be caused by the microstructure or the specimen orientation. Sulfur and magnesium came from the immersion solution (ASTM seawater), which must have remained on the coupons. The oxygen concentration of BM 6061, HAZ 6061, and TMAZ 6061 is higher than BM 5086, HAZ 5086, and TMAZ 5086, respectively. It indicates that AA6061 easily corroded than AA 5086, which agrees with immersion results in Fig. 8. Comparing oxygen concentration of point 4 and 5 in Fig. 11d, higher point 4 has more oxygen concentration then lower point 5. This is because corroded production (point 4) accumulated on sample surface (point 5). Low oxygen concentration appears in HAZ 5086 and HAZ 6061. Low oxygen concentration means high corrosion resistance, which also agrees with $I_{corr}$ results in Table 2 and 3. Pits with different sizes were also observed in the HAZ 5086 (Fig. 11b) and HAZ 6061 (Fig. 11f) regions. The pits around the constituent particles likely formed as a result of cathodic reduction [27]. Cathodic reduction on the constituent particles can increase the alkalinity in the surrounding solution, leading to the dissolution of the aluminum matrix.

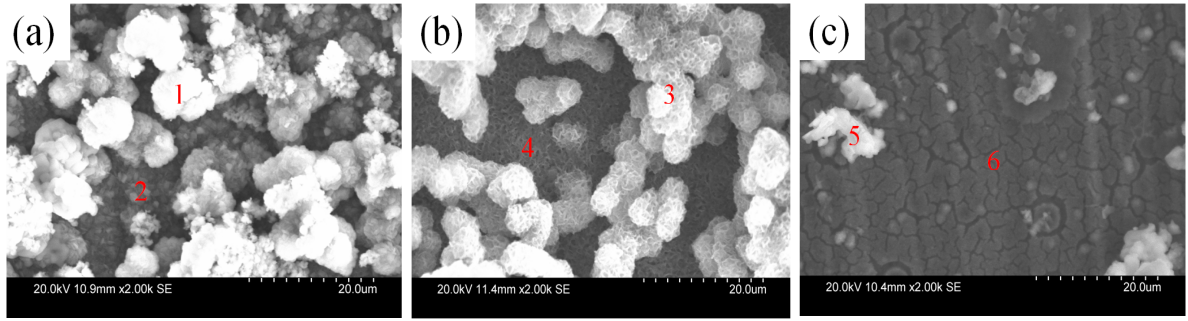

Fig. 12. SEM of NZ of FSW AA5086-AA6061 after 120 days immersion in (a) 3.15 wt% NaCl, (b) ASTM seawater, and (c) 0.5 M $Na_2SO_4$ solution.

Table 4. EDXA quantification results of six points in Fig. 12.

| Element [at. %] | S | O | Na | Al | Cl | Mg |
|---|---|---|---|---|---|---|
| Spectrum 1 | 0.00 | 78.34 | 1.09 | 19.55 | 1.03 | 0.00 |
| Spectrum 2 | 0.00 | 66.12 | 0.80 | 29.98 | 3.10 | 0.00 |
| Spectrum 3 | 2.59 | 66.89 | 4.27 | 10.84 | 5.81 | 9.61 |
| Spectrum 4 | 4.30 | 54.01 | 0.00 | 21.87 | 12.56 | 7.26 |
| Spectrum 5 | 10.34 | 62.70 | 22.88 | 4.08 | 0.00 | 0.00 |
| Spectrum 6 | 1.42 | 41.30 | 1.55 | 55.73 | 0.00 | 0.00 |

Fig. 12 shows SEM of NZ of FSW AA5086-AA6061 after 120 days immersion in 3.15 wt% NaCl, ASTM seawater, and 0.5 M $Na_2SO_4$ solution. The EDXA analysis depicted in Table 4 revealed corrosion product high in oxygen and aluminum. According to Raman analysis (Fig. 10), the corrosion products on the surface of the coupon immersed in 3.15 wt% NaCl and ASTM seawater was $Al(OH)_3$. A denser layer of corrosion products appeared to cover the coupon exposed in the NaCl solution as compared to the coupon exposed in the ASTM seawater. The oxygen concentration of the two locations shown in Fig. 12a (specimen previously exposed to NaCl solution) was also higher than in Fig. 12b (specimen previously exposed to the ASTM solution). As shown in Fig. 12c, cracking on the surface of the specimen exposed to the 0.5 M $Na_2SO_4$ solution was also found. The cracks are likely to be in the aluminum oxide corrosion layer.

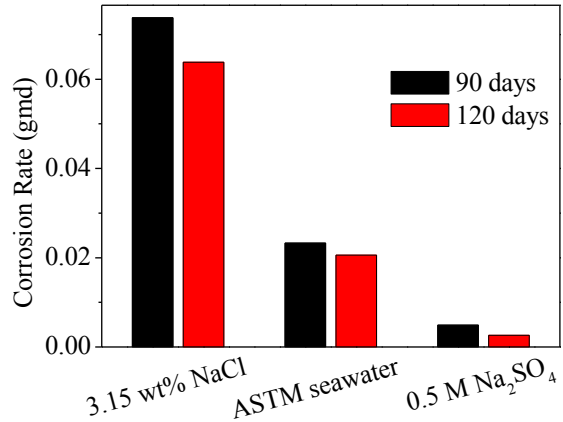

Fig. 13. Corrosion rate of specimens after 90 and 120 days immersion in 3.15 wt% NaCl, ASTM seawater, and 0.5 M $Na_2SO_4$ solution for FSW AA5086-AA6061.

The results of corrosion rate of FSW AA5086-AA6061 by weight measurement are shown in Fig. 13. The corrosion rate of FSW AA5086-AA6061 decreased from 90 days to 120 days in three solutions. The corrosion rate of FSW AA5086-AA6061 from highest to lowest for three solutions was as follows: 3.15 wt% NaCl > ASTM seawater > 0.5 M $Na_2SO_4$. The maximum value of corrosion rate was obtained from FSW AA5086-AA6061 in 3.15 wt% NaCl solution after 90 days immersion. The corrosion rates of FSW AA5086-AA6061 after 90 days immersion are higher than those from the 120 days immersion in three solutions. Therefore, the buildup of corrosion products slowed down the corrosion of the Al samples.

*4 Conclusions*

The mechanical properties and corrosion behavior of FSW AA5086 and AA6061 aluminum alloys were studied in this paper. The minimum hardness of 88 HV was obtained in the HAZ 6061, and tensile specimens also failed in the HAZ 6061. The ductility of FSW AA5086-AA6061 increased in the longitudinal direction (that contained the NZ), but it decreased in the transverse direction that cut through all weld zones. The $E_{corr}$ values of the weld zones shifted to more positive potentials from BM 5086 and BM 6061. The current $I_{corr}$ values for the various weld zones decreased from that of the BM 5086 (1.70 µA/cm$^2$) and BM 6061 (2.44 µA/cm$^2$). The minimum value of $I_{corr}$ appeared in the HAZ-AA5086 (0.54 µA/cm$^2$) region. During cathodic polarization in aerated 3.15 wt% NaCl, the weld zones on the AA6061 side showed diffusion-limited oxygen reduction behavior; whereas, the weld zones on the AA5086 side show Tafel behavior. HAZ 5086 and HAZ 6061 appear much better corrosion resistance than other regions. AA 5086 side of coupon shows better corrosion resistance than AA 6061 side. Raman results revealed Al(OH)$_3$ (Bayerite) as the main corrosion product on FSW AA5086-AA6061 coupons immersed in the three solutions. Corrosion rate of 90 days immersion is higher than 120 days immersion in three solution due to corroded production accumulated on the surface of coupons resulting into corrosion rate decrease.

*Acknowledgements*

The authors are grateful for the financial support from the Office of the Under Secretary of Defense for the project entitled **"Correlation of Field and Laboratory Studies on the Corrosion of Various Alloys in a Multitude of Hawaii Micro-Climates"** (U.S. Air Force Academy, Contract no.: FA7000-10-2-0010). The authors are particularly grateful to Mr. Daniel Dunmire, Director, Corrosion Policy and Oversight, Office of the Under Secretary of Defense.

**References**

[1] W.M. Thomas, E.D. Nicholas, J.C. Needham, M.G. Murch, P. Temple-Smith, C.J. Dawes, Friction welding, in, Google Patents, 1995.

[2] A.R. Ilkhichi, R. Soufi, G. Hussain, R.V. Barenji, A. Heidarzadeh, Establishing mathematical models to predict grain size and hardness of the friction stir-welded AA 7020 aluminum alloy joints, Metallurgical and Materials Transactions B 46 (2015) 357-365.

[3] A. Heidarzadeh, R.V. Barenji, M. Esmaily, A.R. Ilkhichi, Tensile properties of friction stir welds of AA 7020 aluminum alloy, Transactions of the Indian Institute of Metals (2015) 1-11.

[4] P. Cavaliere, A. De Santis, F. Panella, A. Squillace, Effect of welding parameters on mechanical and microstructural properties of dissimilar AA6082–AA2024 joints produced by friction stir welding, Materials & Design 30 (2009) 609-616.

[5] T. Chen, Process parameters study on FSW joint of dissimilar metals for aluminum–steel, Journal of materials science 44 (2009) 2573-2580.

[6] R. Prado, L. Murr, D. Shindo, K. Soto, Tool wear in the friction-stir welding of aluminum alloy 6061+ 20% $Al_2O_3$: a preliminary study, Scripta Materialia 45 (2001) 75-80.

[7] J.-Q. Su, T. Nelson, R. Mishra, M. Mahoney, Microstructural investigation of friction stir welded 7050-T651 aluminium, Acta Materialia 51 (2003) 713-729.

[8] K. Jata, K. Sankaran, J. Ruschau, Friction-stir welding effects on microstructure and fatigue of aluminum alloy 7050-T7451, Metallurgical and materials transactions A 31 (2000) 2181-2192.

[9] M.A. Sutton, B. Yang, A.P. Reynolds, R. Taylor, Microstructural studies of friction stir welds in 2024-T3 aluminum, Materials science and engineering: A 323 (2002) 160-166.

[10] W. Xu, J. Liu, G. Luan, C. Dong, Temperature evolution, microstructure and mechanical properties of friction stir welded thick 2219-O aluminum alloy joints, Materials & Design 30 (2009) 1886-1893.

[11] W. Xu, J. Liu, Microstructure and pitting corrosion of friction stir welded joints in 2219-O aluminum alloy thick plate, Corrosion Science 51 (2009) 2743-2751.

[12] S. Tutunchilar, M.B. Givi, M. Haghpanahi, P. Asadi, Eutectic Al–Si piston alloy surface transformed to modified hypereutectic alloy via FSP, Materials Science and Engineering: A 534 (2012) 557-567.

[13] P. Cavaliere, E. Cerri, A. Squillace, Mechanical response of 2024-7075 aluminium alloys joined by Friction Stir Welding, Journal of Materials Science 40 (2005) 3669-3676.

[14] S.A. Khodir, T. Shibayanagi, Friction stir welding of dissimilar AA2024 and AA7075 aluminum alloys, Materials Science and Engineering: B 148 (2008) 82-87.

[15] M.B. Prime, T. Gnäupel-Herold, J.A. Baumann, R.J. Lederich, D.M. Bowden, R.J. Sebring, Residual stress measurements in a thick, dissimilar aluminum alloy friction stir weld, Acta Materialia 54 (2006) 4013-4021.

[16] J. Ouyang, R. Kovacevic, Material flow and microstructure in the friction stir butt welds of the same and dissimilar aluminum alloys, Journal of Materials Engineering and Performance 11 (2002) 51-63.

[17] I. Shigematsu, Y.-J. Kwon, K. Suzuki, T. Imai, N. Saito, Joining of 5083 and 6061 aluminum alloys by friction stir welding, Journal of Materials Science Letters 22 (2003) 353-356.

[18] H. Jamshidi Aval, S. Serajzadeh, A. Kokabi, A. Loureiro, Effect of tool geometry on mechanical and microstructural behaviours in dissimilar friction stir welding of AA 5086-AA 6061, Science and Technology of Welding and Joining 16 (2011) 597-604.

[19] M. ILANGOVAN, S.R. BOOPATHY, V. BALASUBRAMANIAN, Microstructure and tensile properties of friction stir welded dissimilar AA6061–AA5086 aluminium alloy joints, Transactions of Nonferrous Metals Society of China 25 (2015) 1080-1090.

[20] Z. Chen, S. Li, L.H. Hihara, Microstructure, mechanical properties and corrosion of friction stir welded 6061 Aluminum Alloy, arXiv preprint arXiv:1511.05507 (2015).

[21] Z. Chen, Microstructure characterization, mechanical properties, and corrosion behaviors of friction stir welded AA 5086 and AA6061, M.Sc. Thesis, Department of Mechanical Engineering, University of Hawaii at Manoa, Honolulu, HI, USA, 2014.

[22] Z. Chen, S. Li, K. Liu, L.H. Hihara, A study on the mechanical property and corrosion sensitivity of an AA5086 friction stir welded joint, arXiv preprint arXiv:1511.04990 (2015).

[23] Z. Hu, X. Wang, S. Yuan, Quantitative investigation of the tensile plastic deformation characteristic and microstructure for friction stir welded 2024 aluminum alloy, Materials Characterization 73 (2012) 114-123.

[24] S. Li, Marine atmospheric corrosion initiation and corrosion products characterization, University of Hawaii Manoa, 205 (2010).

[25] S. Li, L.H. Hihara, In situ Raman spectroscopic study of NaCl particle-induced marine atmospheric corrosion of carbon steel, Journal of the Electrochemical Society 159 (2012) C147-C154.

[26] S. Li, L.H. Hihara, In situ Raman spectroscopic identification of rust formation in Evans' droplet experiments, Electrochemistry Communications 18 (2012) 48-50.

[27] M. Jariyaboon, A.J. Davenport, R. Ambat, B.J. Connolly, S.W. Williams, D.A. Price, The effect of welding parameters on the corrosion behaviour of friction stir welded AA2024–T351, Corrosion Science 49 (2007) 877-909.